\documentclass[11pt,a4paper,fleqn]{article}
\usepackage[utf8]{inputenc}
\usepackage{authblk}
\usepackage{geometry}
\usepackage{xcolor}
\usepackage[longnamesfirst]{natbib}
\usepackage{graphicx}
\usepackage{pdflscape}  
\usepackage{appendix}
\usepackage{threeparttable}
\usepackage{booktabs}

\usepackage{amssymb}
\usepackage{amsbsy}
\usepackage{bm}
\usepackage{amsmath}
\usepackage{amsthm}
\usepackage{graphicx}
\usepackage{mathtools}
\usepackage{rotating}
\usepackage{setspace}
\newtheorem{exa}{Example}

\usepackage{color, colortbl}
\definecolor{ashgrey}{rgb}{0.7, 0.75, 0.71}
\definecolor{columbiablue}{rgb}{0.61, 0.87, 1.0}
\definecolor{coral}{rgb}{1.0, 0.5, 0.31}

\definecolor{colBVAR}{HTML}{bababa}
\definecolor{colBART}{HTML}{d7191c}
\definecolor{colmixBART}{HTML}{fdae61}
\definecolor{colerrorBART}{HTML}{abd9e9}
\definecolor{colfullBART}{HTML}{2c7bb6}

\definecolor{colcons}{HTML}{e31a1c}
\definecolor{colSV}{HTML}{a6cee3}
\definecolor{colhBART}{HTML}{1f78b4}

\usepackage{enumitem}
\newlist{steps}{enumerate}{1}
\setlist[steps,1]{label = Step \arabic*:}

\usepackage{adjustbox}
\usepackage{dcolumn}
\newcolumntype{d}[1]{D..{#1}} 

\usepackage{tikz}
\usetikzlibrary{shapes,arrows}	
\usepackage{tkz-berge}
\usetikzlibrary{bayesnet}

\usepackage{caption}
\usepackage{subcaption}
\definecolor{nblue}{HTML}{000660}
\usepackage[colorlinks=true,urlcolor=nblue,linkcolor=nblue,citecolor=nblue]{hyperref}

\newcommand*{\myeqref}[2][Eq.~]{%
  \hyperref[{#2}]{#1(\ref*{#2})}%
}
\def\equationautorefname#1#2\null{%
  Eq.#1(#2\null)%
}

\begin{document}

\title{\textbf{Inference in Bayesian Additive Vector Autoregressive Tree Models}\thanks{
We would like to thank two anonymous reviewers, an associated editor as well as the handling editor for numerous comments that greatly improved the paper. Florian Huber gratefully acknowledges funding from the Austrian Science Fund (FWF): ZK 35. }}

\author[a]{Florian \textsc{Huber}}
\author[b]{Luca \textsc{Rossini}}
\affil[a]{\textit{University of Salzburg}}
\affil[b]{\textit{Queen Mary University of London}}
\date{\today}

\maketitle\thispagestyle{empty}\normalsize\vspace*{-2em}\small\linespread{1.5}
\begin{center}
\begin{minipage}{0.8\textwidth}
\noindent\small Vector autoregressive (VAR) models assume linearity between the endogenous variables and their lags. This  assumption might be overly restrictive and could have a deleterious impact on forecasting accuracy. As a solution, we propose  combining VAR with Bayesian additive regression tree (BART) models. The resulting Bayesian additive vector autoregressive tree (BAVART) model is capable of capturing arbitrary non-linear relations between the endogenous variables and the covariates without much input from the researcher. Since controlling for heteroscedasticity is key for producing precise density forecasts, our model allows for stochastic volatility in the errors. We apply our model to two datasets. The first application shows that the BAVART model yields highly competitive forecasts of the US term structure of interest rates. In a second application, we estimate our model using a moderately sized Eurozone dataset to investigate the dynamic effects of uncertainty on the economy. 

\end{minipage}
\end{center}

\normalsize\newpage
\section{Introduction}
\label{sec_Intro}
In macroeconomics and finance, most models commonly employed to study the transmission of economic shocks or produce predictions assume linearity in their parameters and are fully parametric. One prominent example is the vector autoregressive (VAR) model that is extensively used in central banks and academia \citep[see][]{Sims1980,Doan1984,Litterman1986,Sims1998}. In normal times, with macroeconomic relations remaining stable, this linearity assumption might describe the data well. In turbulent times, however, key transmission channels often change and more flexibility could be necessary. Ignoring such changes or failing to effectively control for outliers could translate into weak out-of-sample forecasts and potentially has adverse effects on the estimation of impulse responses. 

The linearity assumption has been subject to substantial criticism  in the  literature. For instance, \cite{Granger1993} show that macroeconomic and financial quantities depend non-linearily on each other and thus assuming linearity might be overly restrictive. As a solution, researchers increasingly rely on non-linear models which feature time-varying parameters. These models either allow for gradually evolving coefficients or feature a rather low number of structural breaks. All these models have in common that within each point in time,  the relationship between the endogenous and explanatory variables is linear and deciding on the specific law of motion is an important modeling decision. 


Another strand of the literature proposes Bayesian nonparametric time series models in order to relax the linearity assumption. In particular, several recent papers \citep[see, e.g.][]{Bassetti2014,Kalli2018,Billio2019} propose novel methods that assume the transition densities to be nonparametric. These techniques are characterized by featuring an infinite dimensional parameter space that is flexibly adjusted to the complexity of the data at hand.\footnote{For a review on Bayesian nonparametric methods, see \cite{Hjort2010}.} Within the nonparametric paradigm, there has been a number of popular competing approaches that make use of machine learning techniques such as boosting \citep{Freund1997,Friedman2001}, bagging and random forests \citep{Breiman2001}, decision trees and Mondrian forests \citep{Roy2009,Lakshminarayanan2014} and Bayesian additive regression tree (BART) models \citep{Chipman2002, Gramacy2008, Chipman2010, Linero2018}. 

In this paper, our focus is on BART models and applying them to multivariate time series data.  BART has been successfully applied  for dealing with model uncertainty \citep{hernandez2018bayesian}, sparse regression problems \citep{Linero2018}, spatial models  as well as for limited dependent variable models \citep{krueger2020new}. Moreover, in several recent studies, BART has been shown to yield precise forecasts and improve upon several competing models from the statistics and machine learning literature \citep[see][]{kapelner2015prediction, waldmann2016genome, Linero2018, he2019xbart, Pruser2019, Huber2021}. But BART has also been successfully used for carrying out causal effect estimation \citep[see, e.g.,][]{hill2011bayesian, green2012modeling, kern2016assessing, dorie2019automated, hahn2020bayesian}.

The strong empirical performance of BART for forecasting and causal inference gives rise to the main contribution of the  paper.  We aim to bridge the literature on BART  \citep[see][]{Chipman2010} with the literature on VAR models. BART, being a flexible nonparametric regression approach,  allows for unveiling non-linear relations between a set of endogenous  and explanatory variables without needing much input  from the researcher. Intuitively speaking, it models the conditional mean of the regression model by summing over a large number of trees which are, by themselves, constrained through a regularization prior. The resulting individual trees will take a particularly simple form and can thus be classified as ''weak learners''. Each of these simple trees only explains a small fraction of the variation in the response variable while a large sum of them is capable of describing the data extremely well.

It is precisely this intuition on which we build on when we generalize these techniques to a multivariate setting. More precisely, we assume that a  potentially large dimensional vector of endogenous variables is determined by its lagged values. As opposed to VAR models which model this relationship using a linear function, we assume the precise functional form to be unknown. This function is then estimated using BART. The resulting model, labeled the Bayesian Additive Vector Autoregressive tree (BAVART) model, is a highly flexible variant that can be used for forecasting and impulse response analysis. 

Estimation and inference are carried out  using Markov chain Monte Carlo (MCMC) techniques. Since the error covariance matrix is non-diagonal, we propose methods that allow for equation-by-equation estimation of the multivariate model. These techniques imply that model estimation scales well in high dimensions and permits estimation of huge dimensional  models. Another novel feature of our approach is that it allows for flexibly handling heteroscedasticity. We control for time-variation in the error variances by proposing a stochastic volatility specification. This feature is crucial for producing precise density forecasts. To produce higher-order forecasts and impulse response functions (IRFs) we develop techniques that enable us to sample from the predictive distribution and the posterior distribution of the IRFs. This proves to be  another important contribution of the paper which is closely related to the literature on estimating non-linear impulse response functions \citep[see, e.g.,][]{barnichon2018functional, plagborg2019bayesian}.

We illustrate our BAVART model using two empirical applications. The first deals with forecasting the US term structure of interest rates. Since financial data often exhibit non-linear features, our BAVART model might be well suited for dealing with interest rates at differing maturities. We consider two ways of modeling the yield curve. The first models a panel of yields simultaneously along the lines of \cite{carriero2012forecasting} whereas the second approach fits a Nelson-Siegel (NS) model in the spirit of \cite{diebold2006forecasting} but assumes the three NS factors to follow a BAVART model. Within each of these classes, we consider several linear and non-linear competing models.  Since interest not only centers on one-step-ahead predictive distributions but also on multi-step-ahead forecasts, we provide algorithms to simulate from the relevant predictive distributions. The findings show that for one-month-ahead point forecasts, our proposed BAVART model with NS factors yields highly competitive forecasts for bonds with maturities greater than five years. For higher-order forecasts, these accuracy gains become smaller. When the full predictive distribution is considered, jointly modeling the yields without imposing a factor structure yields more precise density predictions. As opposed to point forecasts, we find that using BART on the conditional mean helps in improving density forecasts for maturities below seven years and both one-step-ahead and three-steps-ahead predictions.

In a second application, we apply the BAVART model to macroeconomic data for the Eurozone. Instead of using our model to produce forecasts, we analyze the dynamic effects of uncertainty on the Eurozone economy. As opposed to the existing literature which deals with the macroeconomic effects of uncertainty using non-linear models \citep[see, e.g.,][]{caggiano2014uncertainty, aastveit2017economic, ferrara2018macroeconomic, mumtaz2018changing,alessandri2019financial,cuaresma2020fragility, jackson2020nonlinear,  paccagnini2020asymmetric, caggiano2017uncertainty}, our approach remains agnostic on the precise form of non-linearities and infers these in a data-based manner. This constitutes a substantial advantage since there is considerable uncertainty with respect to selecting the appropriate type of non-linearities in multivariate time series models.

To assess how macroeconomic reactions change in our non-linear and non-parametric framework if interest rates are at the zero lower bound, we simulate dynamic responses under the restriction that short-term interest rates are zero and not allowed to react to uncertainty shocks. Our findings indicate that increases in economic uncertainty have negative effects on real activity. Specifically, we observe increases in unemployment rates, declining consumption levels, and a drop in prices. By contrast, financial market quantities display adverse reactions with declines in stock prices and decreasing short-term interest rates. These findings are in general consistent with established findings in the literature and thus show that our BAVART model can be used to carry out meaningful structural inference.

The remainder of this paper is organized as follows. Section \ref{sec_BARTModel} discusses the BART model   in the context of the homoscedastic regression model while Section \ref{sec:BAVART} extends this method to the VAR case and proposes the BAVART specification and how we control for heteroscedasticity. Section \ref{sec: forecasting}  presents the results of the term structure forecasting exercise while Section \ref{sec: IRF_EA}  investigates the relationship between uncertainty and macroeconomic dynamics in the Eurozone. Finally, the last section summarizes our findings and concludes the paper.

%
\section{Bayesian Additive  Regression Tree Models}
\label{sec_BARTModel}
In this section, we briefly review BART.\footnote{For an extensive introduction to BART models, see \cite{Hill2020}.}  Let $\bm y = (y_1, \dots, y_T)'$ denote the $T-$dimensional response vector  and $\bm X =(\bm x_1, \dots, \bm x_T)'$ be a $T \times K$-dimensional matrix of exogenous variables with $\bm x_t$ being the $K$ covariates in time $t$. We assume that $\bm y$ depends on $\bm X$ through a potentially non-linear function $f: \mathbb{R}^{T \times K} \to \mathbb{R}^T$ as follows:
\begin{equation*}
\bm y = f(\bm X) + \bm \varepsilon, \quad \bm \varepsilon \sim \mathcal{N}(\bm 0_T, \sigma^2\bm I_T),
\end{equation*}
where $\sigma^2$ denotes the error variance and the function $f$ is generally not known.  BART approximates the function $f$ by summing over $N$ (which is a large number) regression trees:
\begin{equation}
f(\bm{X}) \approx \sum_{j=1}^{N} g(\bm{X} | \mathcal{T}_j,\bm{m}_j).\label{eq_Tree}
\end{equation}
In \myeqref{eq_Tree}, the function $g(\bm X | \mathcal{T}_j,\bm{m}_j)$ corresponds to a single tree model with $\mathcal{T}_j$ denoting the tree structure associated with the $j^{th}$ binary tree  and $\bm{m}_j = (\mu_{j1},\ldots,\mu_{{jb}_{j}})'$ is the vector of terminal node parameters associated with $\mathcal{T}_j$ and $b_j$ are the leaves of the $j^{th}$ tree.  In what follows, and in consistency with \cite{Chipman2010}, we set $N=250$ in all our empirical applications. 

These binary trees are constructed by considering splitting rules of the form $\lbrace \bm X \in \mathcal{A}_{jk}\rbrace$ or $\lbrace \bm X \not\in \mathcal{A}_{jk}\rbrace$ with $\mathcal{A}_{jk}$ being a partition set.  These rules typically depend on selected columns of $\bm X$, denoted as $\bm X_{\bullet j}~(j=1, \dots, K)$, and a threshold $c$. The set  $\mathcal{A}_{jk}$ is then defined by splitting the predictor space according to $\lbrace \bm X_{\bullet j} \le c \rbrace$ or $\lbrace \bm X_{\bullet j} > c \rbrace$.

The step function $g$ is constant over the elements of  $\mathcal{A}_{jk}$:
\begin{equation*}
g(\bm{X}| T_j,\bm{m}_j) = \mu_{jk}, \quad \text{ if } \bm{X} \in \mathcal{A}_{jk}, \quad k=1,\ldots,b_j.
\end{equation*}
Hence, the set $\mathcal{A}_{jk}$ defines a tree-specific unique partition of the covariate space such that the function $g$ returns a specific value $\mu_{jk}$ for specific values of $\bm x_t$.

To avoid overfitting, the trees are encouraged to be small (i.e. take a particularly simple form) and the terminal node parameters to be shrunk to zero. If the first tree, $g(\bm{x}|\mathcal{T}_1,\bm{m}_1)$, is a weak learner and fitted in a reasonable way, the corresponding tree structure will be very simple and elements in $\bm{m}_1$ will be pushed towards zero. This implies that the first tree will explain a small fraction of the variation in $\bm y$. Subtracting $g(\bm X| \mathcal{T}_1, \bm{m}_1)$ from $\bm y$ yields a new conditional model with transformed $\tilde{\bm y} = \bm y - g(\bm X| \mathcal{T}_1, \bm{m}_1)$ and then the next tree will be fitted with $\bm y$ being replaced by $\tilde{\bm y}$. This procedure is repeated for a sufficiently large number $N$ of trees until the fit of the additive model becomes reasonably good.

We illustrate BART using a simple example. In this simple example, we first focus on a single regression tree model (i.e. $N=1$). The case of several regression trees is considered afterwards.

\begin{figure}[h!]
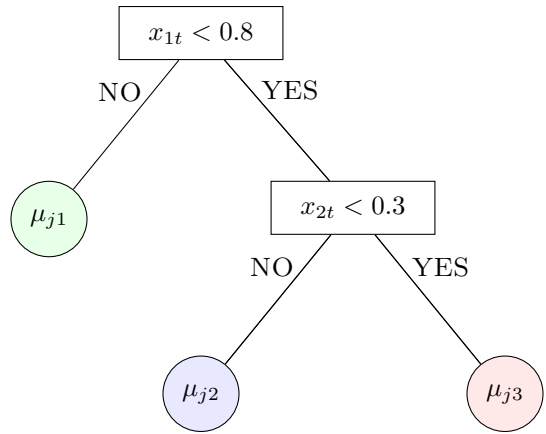

  \begin{center}
    \begin{tabular}{cc}
    \tikz{ %
        \node[latent,rectangle] (first) {$\quad x_{1t} <0.8 \quad$} ; %
        \node[latent, circle, below=of first, xshift = -2cm, yshift = -0.6cm, fill=green!9, minimum size=10mm] (second) {$\mu_{j1}$};  %
        \node[latent, rectangle, below=of first, xshift = 2cm, yshift = -0.6cm] (third) {$\quad x_{2t}<0.3 \quad$}  edge [-] (first);%
        \node[latent, circle, below=of third, xshift = -2cm, yshift = -0.6cm, fill=blue!9,minimum size=10mm] (forth) {$\mu_{j2}$}  edge [-] (third);%
        \node[latent, circle, below=of third, xshift = 2cm, yshift = -0.6cm, fill=red!9, minimum size=10mm] (fifth) {$\mu_{j3}$}  edge [-] (third);%
        \path[-]    (first) edge[-]     node[swap, near start,left]   {\small{\textsc{NO}}} (second)
          edge[-]     node[swap, near start,right]   {\small{\textsc{YES}}} (third)
          (third) edge[-]     node[swap, near start,left]   {\small{\textsc{NO}}} (forth)
          edge[-]     node[swap, near start,right]   {\small{\textsc{YES}}} (fifth)
           }
    \end{tabular}
  \end{center}
\caption{Example of binary regression tree, $g(\bm{x}| \mathcal{T}_j,\bm{m}_j)$, with internal nodes labeled by their splitting rules and leaf nodes labeled with the corresponding parameters $\mu_{jk}$, with $j=1$ and $k=1,2,3$. \label{fig_singletree}}
\end{figure}

\begin{exa}
Consider the regression tree $g(\bm{X}| \mathcal{T}_j,\bm{m}_j)$ in Figure \ref{fig_singletree}. We include two covariates in $\bm x_t = (x_{1t}, x_{2t})$. In the figure, each rectangle refers to a splitting rule and we henceforth label this a node.  At the top node (also called root node or simply root), we have the condition that asks whether $x_{1t} <0.8$. If this  condition is not true, then we arrive at the terminal node, $\mu_{j1}$ and set $\mathbb E(y_t) = \mu_{j1}$. By contrast, if the condition is fulfilled we move to the next node. The corresponding splitting rule  states that if $x_{2t} \ge 0.3$, we reach a terminal node with $\mathbb E(y_t) = \mu_{j2}$ whereas if $x_{2t} < 0.3$, we set $\mathbb E(y_t) = \mu_{j3}$. Using a single tree thus yields three possible values for the conditional mean, $\mu_{j1}, \mu_{j2}$ and $\mu_{j3}$ and implies abrupt shifts between them. If $y_t$ is evolving smoothly over time, using a single tree would thus lead to a rather simplistic conditional mean structure.
\end{exa}

\begin{figure}[h!]
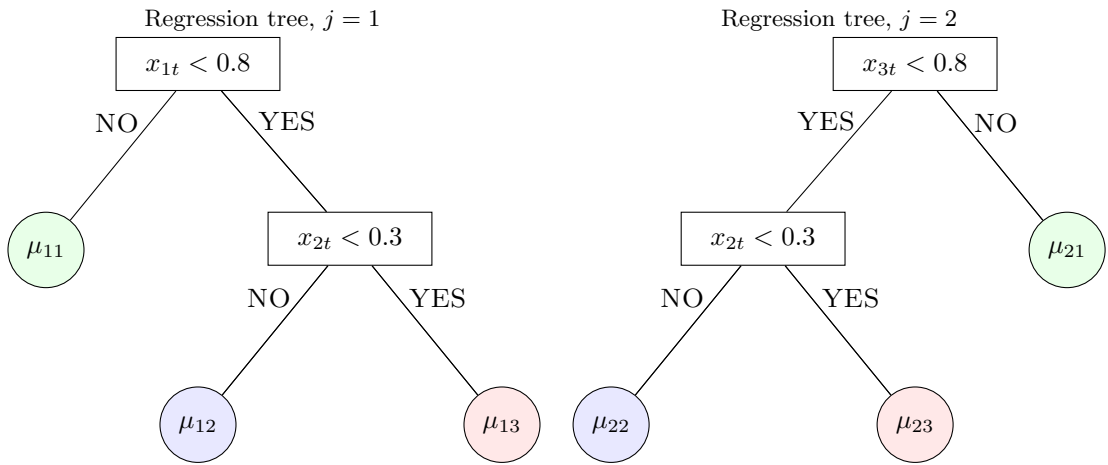

  \begin{center}
    \begin{tabular}{cc}
    \footnotesize{Regression tree, $j=1$} &  \footnotesize{Regression tree, $j=2$} \\
    \hspace{5pt}  
    \tikz{ %
        \node[latent,rectangle] (first) {$\quad x_{1t} <0.8 \quad$} ; %
        \node[latent, circle, below=of first, xshift = -2cm, yshift = -0.6cm, fill=green!9, minimum size=10mm] (second) {$\mu_{11}$};  %
        \node[latent, rectangle, below=of first, xshift = 2cm, yshift = -0.6cm] (third) {$\quad x_{2t}<0.3 \quad$}  edge [-] (first);%
        \node[latent, circle, below=of third, xshift = -2cm, yshift = -0.6cm, fill=blue!9,minimum size=10mm] (forth) {$\mu_{12}$}  edge [-] (third);%
        \node[latent, circle, below=of third, xshift = 2cm, yshift = -0.6cm, fill=red!9, minimum size=10mm] (fifth) {$\mu_{13}$}  edge [-] (third);%
        \path[-]    (first) edge[-]     node[swap, near start,left]   {\small{\textsc{NO}}} (second)
          edge[-]     node[swap, near start,right]   {\small{\textsc{YES}}} (third)
          (third) edge[-]     node[swap, near start,left]   {\small{\textsc{NO}}} (forth)
          edge[-]     node[swap, near start,right]   {\small{\textsc{YES}}} (fifth)
           } &
               \tikz{ %
        \node[latent,rectangle] (first) {$\quad x_{3t} <0.8 \quad$} ; %
        \node[latent, rectangle, below=of first, xshift = -2cm, yshift = -0.6cm] (third) {$\quad x_{2t}<0.3 \quad$};  %
        \node[latent, circle, below=of first, xshift = 2cm, yshift = -0.6cm, minimum size=10mm, fill=green!9] (second)  {$\mu_{21}$} edge [-] (first);%
        \node[latent, circle, below=of third, xshift = -2cm, yshift = -0.6cm, fill=blue!9,minimum size=10mm] (forth) {$\mu_{22}$}  edge [-] (third);%
        \node[latent, circle, below=of third, xshift = 2cm, yshift = -0.6cm, fill=red!9, minimum size=10mm] (fifth) {$\mu_{23}$}  edge [-] (third);%
        \path[-]    (first) edge[-]     node[swap, near start,right]   {\small{\textsc{NO}}} (second)
          edge[-]     node[swap, near start,left]   {\small{\textsc{YES}}} (third)
          (third) edge[-]     node[swap, near start,left]   {\small{\textsc{NO}}} (forth)
          edge[-]     node[swap, near start,right]   {\small{\textsc{YES}}} (fifth)
           }
    \end{tabular}
  \end{center}
\caption{Example of a sum of regression tree, $g(\bm{x}|\mathcal{T}_j,\bm{m}_j)$, with internal nodes labeled by their splitting rules and leaf nodes labeled with the corresponding parameters $\mu_{jk}$, where $j=1,2$ and $k=1,2,3$. \label{fig_sumtree}}
\end{figure}

\begin{exa}
Since the  tree model in the first example is too simplistic, our second example deals with a more flexible conditional mean structure. Let us consider a sum of regression trees with $m=2$ trees and $3$ covariates, depicted in Figure \ref{fig_sumtree}. Within a given tree, the same intuition as in Example 1 applies. However, since we now use two trees we gain more flexibility in modeling the conditional mean of $y_t$. This is because the conditional mean at a given point in time is equal to the sum of the terminal node parameters for the two trees and the combination of the two parameters depends on the specific set of decision rules across both trees.  This allows for a substantially richer mean structure and increased flexibility as opposed to a single tree model.
\end{exa}

The second example shows how flexibility is increased by adding more trees and illustrates how BART handles non-linearities in a  flexible manner.  In particular, each regression tree is a simple step-wise function and when we sum over the different regression trees, we gain flexibility. The resulting additive model essentially allows for approximating non-linearities without prior assumptions on the specific form of the non-linearities.


\section{BART models for multivariate time series analysis}
\label{sec:BAVART}


\subsection{Bayesian Additive Vector Autoregressive Tree Models}
In this section we generalize the model outlined in the previous section to the multivariate case. Consider a $M$-dimensional vector of endogenous variables $\bm y_t = (y_{1t}, \dots, y_{Mt})'$. Stacking the rows yields a $T \times M$ matrix $\bm Y =  (\bm y_1, \dots, \bm y_T)'$. We assume that $\bm Y$ depends on a $T \times K$ matrix $\bm X = (\bm X_1, \dots, \bm X_T)'$ with each $\bm X_t = (\bm y'_{t-1}, \dots, \bm y'_{t-P})'$ being a $K(=PM)$-dimensional vector of lagged endogenous variables. The BAVART model is then given by:
\begin{equation*}
    \bm y_t = F(\bm X_t) + \bm \varepsilon_t, 
\end{equation*}
or in terms of full-data matrices:
\begin{equation}
\bm Y = F(\bm X) + \bm \varepsilon,
\label{BAVART_stacked}
\end{equation}
with $\bm \varepsilon = (\bm \varepsilon_1, \dots, \bm \varepsilon_T)'$ denoting a $T \times M$ matrix of shocks with typical $t^{th}$ row $\bm \varepsilon_t \sim \mathcal{N}(\bm 0_M, \bm \Sigma)$. For the moment, we assume that the error variance-covariance matrix $\bm \Sigma$ is time-invariant. 

In \myeqref{BAVART_stacked},  $F$ is defined in terms of equation-specific functions $f_j(\bm{X})$:
\begin{equation*}
F(\bm X) = (f_1(\bm X), \dots, f_M (\bm X))'. 
\end{equation*}
Similarly to the standard BART specification,  we approximate each  $f_j~(j=1, \dots, M)$ through a sum of $N$ regression trees:
\begin{equation*}
f_j(\bm X) \approx \sum_{k=1}^N g_{jk}(\bm X | \mathcal{T}_{jk},\bm{m}_{jk}).
\end{equation*}
Here, $g_{jk} ~(k=1,\dots, N)$ denotes an equation-specific step function with arguments $\mathcal{T}_{jk}$ and $\bm m_{jk}$. As before, the individual tree structures $\mathcal{T}_{jk}$ are associated with a $b_{jk}$-dimensional vector  $\bm{m}_{jk} = (\mu_{jk, 1}, \dots, \mu_{jk, b_{jk}})'$ of {terminal}  node coefficients associated with $b_{jk}$ denoting the  number of leaves per tree in equation $j$. Notice that both the tree structures and the terminal node parameters are now specific to equation $j$. The main difference to the model illustrated in Section \ref{sec_BARTModel} is that we approximate $M$ different functions $f_j$. This implies that if certain elements in $\bm y_t$ depend linearly on $\bm X_t$, then our flexible multivariate specification can pick this up. 

In what follows, we will estimate the BAVART model by exploiting its structural form. Using the structural form of the VAR to speed up computation has been done in several recent papers \citep[see, e.g.,][]{Carriero2019, huber2020inducing}. This implies that \myeqref{BAVART_stacked} can be written as follows:
\begin{equation}
\bm Y = {F}(\bm X) + \bm \epsilon \bm A'_0 ,  \label{eq:strucform}
\end{equation}
whereby  $\bm A_0$ denotes a $M \times M$-dimensional lower triangular matrix with $\text{diag}(\bm A_0) = (1, \dots, 1)'$ and $\bm \epsilon = (\bm \epsilon_1, \dots, \bm \epsilon_T)'$ is a $T \times M$ matrix of orthogonal shocks with $\bm \epsilon_t \sim \mathcal{N}(\bm 0_M,  \bm H)$.   $\bm H$ denotes a $M \times M$-dimensional diagonal matrix with the variances on its main diagonal.  This implies that $\bm \Sigma = \bm A_0 \bm H \bm A'_0$.

Conditional on $\bm A_0$, this form permits equation-by-equation estimation since the shocks are independent. This leads to enormous computational gains. Notice that the $j^{th}>1$ equation can be written as: 
\begin{equation}
\bm y_{\bullet j} = \sum_{k=1}^N {g}_{j k}(\bm X| \mathcal{T}_{j k}, \bm m_{j k}) + \sum_{l=1}^{j-1} {a}_{jl} \bm \varepsilon_{\bullet l} +  \bm \epsilon_{\bullet j}.
\label{BAVART_equation}
\end{equation}
Here, $\bm y_{\bullet j},  \bm \varepsilon_{\bullet l}$ and $\bm \epsilon_{\bullet j}$ refers to the $j^{th}$ or $l^{th}$ column of $\bm Y, \bm \varepsilon$ and $\bm \epsilon$, respectively.  $a_{jl}$ denotes the $(j,l)^{th}$ element of $\bm A_0$. 

Notice that \myeqref{BAVART_equation} is a generalized additive model that consists of a non-parametric part $\left(\sum_{k=1}^N g_{j k}(\bm X| \mathcal{T}_{j k}, \bm m_{j k})\right)$ and a regression part 
$\left(\sum_{l=1}^{j-1} {a}_{jl} \bm \varepsilon_{\bullet l}\right)$. For $j=1$, the model reduces to a standard BART specification without the regression part.

The key idea behind this formulation is that, for a sufficiently large number of trees, we approximate non-linear relations between $\bm y_t$ and its lags while allowing for linear  relations between the contemporaneous values of $\bm \varepsilon_t$. These linear relations determine the covariances and are of vital importance for the identification of structural shocks.

\subsection{Allowing for Heteroscedasticity}\label{sec:heteroscedasticity}
Up to this point we assumed the error variance to be constant. If the researcher wishes to relax this assumption, several feasible options exist. In a recent paper, \cite{Pratola2019} propose a combination of a standard additive BART model with a multiplicative BART specification to flexibly control for heteroscedasticity. But modeling heteroscedasticity with BART implies that we need some information on how the volatility of economic shocks depends on additional covariates (which potentially differ from $\bm X$). As a simple yet flexible solution we adopt a standard stochastic volatility (SV) model. SV models are frequently used in macroeconomics and finance and have a proven track record for forecasting applications \citep[see, e.g.,][]{clark2011real, clark2015macroeconomic}. 

Our SV specification  assumes that $\bm H$ is time-varying:
\begin{equation*}
    \bm H_t = \text{diag}(e^{h_{1t}}, \dots, e^{h_{Mt}})
\end{equation*}
with the time-varying variances $e^{h_{jt}}$. We assume that the $h_{jt}$'s follow an AR(1) model:
\begin{align}
h_{jt} &=  c_j + \rho_j (h_{jt-1}-c_j) + \sigma_{jh} \nu_{jt},\quad  \nu_{jt} \sim \mathcal{N}(0, 1),\\\quad h_{j0} &\sim \mathcal{N}\left(c_j,  \frac{\sigma^2_{jh}}{1-\rho_j^2}\right).\label{eq: stateSV}
\end{align}
Here, $c_j$ is the unconditional mean, $\rho_j$ is the persistence parameter and $\sigma_{j h}^2$ is the error variance of the log-volatility process. This specification essentially implies that, if $\rho_j$ is close to one, the log-volatilities evolve smoothly over time and tend to be persistent. 

For macroeconomic and financial data, using SV has been shown to greatly increase forecast accuracy. It is moreover worth noting that using SV entails a much more flexible error distribution than the one used in, e.g., \cite{Chipman2010}.  Hence, coupling the BAVART model with a stochastic volatility component yields a model which allows for flexible adjustments of the conditional mean while also being flexible on the error variances. In our empirical work, the models considered feature stochastic volatility of this form.

\subsection{The Prior and Posterior Simulation}\label{sec_prior}
The Bayesian approach calls for the specification of suitable priors over the parameters of the model. Here we mainly follow the different strands of the literature our approach combines. In particular, we focus on the priors associated with the trees $\mathcal{T}_{jk}$ and the terminal node parameters $\bm m_{jk}$ for $j=1,\dots, M$ and $k=1, \dots, N$. We assume that the hyperparameters across priors are the same for each equation. On the covariances we use the Horseshoe prior \citep{carvalho2009handling}.

For each equation $j$, the joint prior structure is given by:
\begin{align*}
p\left((\mathcal{T}_{j1},\bm{m}_{j1}) \dots, (\mathcal{T}_{jN},\bm{m}_{jN}), c_j, \rho_j, \sigma^2_j, \bm a_j\right) &= p\left((\mathcal{T}_{j1},\bm{m}_{j1}) \dots, (\mathcal{T}_{jN},\bm{m}_{jN})\right)~ p(c_j)~  p(\rho_j)~ p(\sigma^2_{j, h})~ p(\bm a_j).
\end{align*}
At this level, the prior implies independence between $p\left((\mathcal{T}_{j1},\bm{m}_{j1}) \dots, (\mathcal{T}_{jN},\bm{m}_{jN})\right)$ and the remaining model parameters.  \cite{Chipman2010} further introduce the following factorization:
\begin{align*}
p\left((T_{j1},\bm m_{j1}), \dots, (T_{jN},\bm m_{jN})\right) &= \prod_{k=1}^N p(\bm m_{jk}|\mathcal{T}_{jk}) \times p(\mathcal{T}_{jk})\\
p(\bm m_{jk}| \mathcal{T}_{jk}) &= \prod_{q=1}^{b_{jq}} p(\mu_{jk, q}|\mathcal{T}_{jk}),
\end{align*}
implying that the tree components $(\mathcal{T}_{jk},\bm{m}_{jk})$ are independent of each other and of the remaining parameters while the prior on $\bm{m}_{jk}$ depends on the tree structure. This prior structure allows for integrating out $\bm m_{jk}$  from the posterior of the trees and thus greatly simplifies computation. 

The prior on the tree structure is specified along the lines suggested in \cite{Chipman1998} and \cite{Chipman2010}. Specifically, instead of constructing a prior on the trees directly we construct a tree generating stochastic process that consists of three steps to grow trees.  Let $s=0$ be the first iteration of this tree generating process. Then the following steps for constructing trees are used:
\begin{enumerate}
\item[] Start with the trivial tree that consists of a single terminal node (i.e. its root), which will be denoted by $\eta_{jk}$ for each equation $j$ and $k=1,\ldots,N$.

\item Split the terminal node $\eta_{jk}$ with probability 
\begin{equation}
p_{SPLIT}(\eta_{jk}, \mathcal{T}^{(s)}_{jk}) = \alpha (1+d)^{-\beta}, \quad \alpha \in (0, 1),\quad \beta \ge 0,\quad  d=0, 1,2,\ldots
\end{equation}
where $d$ denotes the depth of the tree, and $(\alpha,\beta)$ are prior hyperparameters.  These parameters are set as follows. In detail, a node at depth $d$ spawns children with $p_{SPLIT}(\eta_{jk}, \mathcal{T}^{(s)}_{jk})$ and as the tree grows, $d$ increases while the prior probability decreases, making it less likely that a node spawns children. This implies that the probability that a given node becomes a bottom node increases and thus a penalty on tree complexity is introduced. We set $\alpha = 0.95$ and $\beta = 2$, implying that the probability that a given node is non-terminal decreases quadratically  if the trees become more complicated (i.e. for increasing levels of $d$).

\item If the current node is split,  we introduce a splitting rule $\kappa_{jk}$ drawn from the distribution function $p_{RULE}(\kappa_{jk}| \eta_{jk}, \mathcal{T}^{(s)}_{jk})$ and use this rule to spawn a left and right children node. In particular, the  rule $p_{RULE}(\kappa_{jk}| \eta_{jk}, \mathcal{T}^{(s)}_{jk})$ is set such that a given ''splitting" covariate $\bm X_{\bullet j}$ is chosen with probability $1/K$. As described above, conditional on a chosen covariate, one needs to estimate a threshold. The prior on this threshold is, in the absence of substantial information, assumed to be uniformly distributed over the range of $\bm X_{\bullet j}$.  

\item Once we obtain  all terminal nodes (i.e. no node is split anymore), we will label this new tree $\mathcal{T}_{jk}^{(s+1)}$ and  return to step (2).
\end{enumerate}

On  the terminal node parameter $\mu_{jk,q}$, we use a  conjugate Gaussian prior distribution $\mathcal{N}(0,\sigma^2_{\mu})$, where $\sigma^2_\mu$ is set as follows:
\begin{equation*}
    \sigma^2_\mu = \frac{\mathfrak{R}_j}{2 \tilde{s} \sqrt{N}},
\end{equation*}
with $\mathfrak{R}_j$ denoting the range of the endogenous variable in equation $j$ and $\tilde{s}$ denotes the number of prior standard deviations. This hyperparameter is set equal to $2$, implying that $\mu_{jk, q}$ will place around $95$ percent prior mass on the range of $\bm y_{\bullet j}$. One key property of this prior is that it increases with $\mathfrak{R}_j$. This implies that if outliers in $\bm Y$ are observed, the range increases and the prior becomes looser. By contrast, it decreases in the number of trees, implying that if $N$ is large, the terminal node parameters associated with a given tree will be strongly pushed towards zero. This is consistent with the notion that we aim to use a composite models of many weak learners as opposed to having a small to moderate number of complex trees.

For the free elements in $\bm A_0$, we introduce a Horseshoe prior on each element of  $\bm{a}_j = (a_{j1},\ldots,a_{jj-1})'$. This prior consists of local scaling parameters $\tau_{jl}$ which are specific to each covariance parameter $a_{jl}$ and a global shrinkage parameter $\lambda$ which pushes all covariances towards zero. The Horseshoe prior on the covariances is given by:
\begin{align*}
a_{jl}|\tau_{jl}, \lambda^2 &\sim \mathcal{N}(0,\tau_{jl}^2 \lambda^2),  \\
\tau_{jl} &\sim \mathcal{C}^{+}(0,1),\\
\lambda &\sim \mathcal{C}^{+}(0,1),
\end{align*}
where $j=1, \dots, M; l=1,\ldots,j-1$ and $\mathcal{C}^{+}$ denotes the half-Cauchy distribution. 

The prior specification on the parameters of the log-volatility equation follows the setup proposed in \cite{Kastner2014}. In details, we use a zero mean Gaussian prior with variance $10^2$ on the unconditional mean $c_j$, a Beta prior on the (transformed) persistence parameter, $\frac{\rho_j+1}{2} \sim \mathcal{B}(25,5)$ and a Gamma prior on the error variance of the log-volatility process $\sigma_{j,h}^2 \sim \mathcal{G}(1/2,1/2)$.

These priors can be combined with the likelihood to yield a joint posterior distribution over the coefficients and latent states in the model. To simulate from this joint posterior distribution we adopt an MCMC algorithm. This algorithm simulates all quantities in an equation-specific manner. Since all steps necessary are standard, we only provide an overview. Appendix \ref{App_Sec_Posterior} provides more detail on sampling the trees. Here it suffices to say that we sample all quantities related to the trees (i.e. $\mathcal{T}_{jk}, \bm m_{jk}$ for all $j, k$) using the algorithm outlined in \cite{Chipman2010}. The latent states and the coefficients associated with the state equation of the log-volatilities are obtained through the efficient algorithm discussed in \cite{Kastner2014}. Conditional on the trees and log-volatilities, the posterior of $\bm a_j$ is multivariate Gaussian and takes a standard form since the resulting conditional model is a linear regression model with heteroscedastic shocks. Finally, the parameters of the Horseshoe prior are simulated using techniques outlined in \cite{makalic2015simple} which involve sampling from inverse Gamma distributions (conditional on introducing auxiliary shrinkage parameters) only.


\section{Forecasting the Term Structure of Interest Rates}
\label{sec: forecasting}
In this section, we illustrate the predictive capabilities of the  BAVART model. After providing an overview of the dataset, the model specification and the design of the forecasting exercise adopted, our focus will be on how well our approach works when applied to US yield curve data. 

\subsection{Data Overview, Model Specification and Design of the Forecasting Exercise}
In this section, we briefly discuss the two datasets adopted. For our forecasting application, we use data on the nominal yield curve which is close to the dataset proposed in \cite{gurkaynak2007us}. These are downloaded from the website of the Federal Reserve  Board (\href{https://www.federalreserve.gov/data/nominal-yield-curve.htm}{https://www.federalreserve.gov/data/nominal-yield-curve.htm}) and range from June 1961 to December 2019. The maturities included are $1, 3, 5, 7, 10$ and $15$ years and we will consider changes in the interest rates. We will then use the period from June 1961 to June 2006 as our initial training sample. This allows us to compute forecast distributions for July 2006. We then move on to expand this estimation period by one month until we reach the end of the full sample. This procedure yields a sequence of 160 predictive distributions which we then evaluate using the observed outcomes.

We  consider two approaches of modeling the term structure of interest rates. Due to its empirical success, we use the three-factor Nelson Siegel (NS) model, originally proposed in \cite{nelson1987parsimonious}, and combined with a VAR state evolution equation on the factors in \cite{diebold2006forecasting}. The NS model assumes that the yield at maturity $\varpi$, $i_t(\varpi)$, depends on three latent factors as follows:
\begin{equation}
    i_t(\varpi) = \mathcal{L}_t + \frac{1 - e^{-\gamma \varpi }}{\gamma \varpi} \mathcal{S}_t +\left(\frac{1 - e^{-\gamma \varpi }}{\gamma \varpi} - e^{- \gamma \varpi}\right) \mathcal{C}_t, \label{eq: NS_factors}
\end{equation}
whereby $\mathcal{L}_t, \mathcal{S}_t$ and $\mathcal{C}_t$ denote a level, slope and curvature factor, respectively. Moreover,  $\gamma$ is a parameter which shapes the factor loadings. Consistent with \cite{diebold2006forecasting} we set $\gamma = 0.0609$. This value maximizes the factor loadings on $\mathcal{C}_t$. The three factors are then obtained on a $t$-by-$t$ basis using OLS estimation. The NS model sets  $\bm y_t = (\mathcal{L}_t, \mathcal{S}_t, \mathcal{C}_t)'$ and assumes it to evolve according to a multivariate dynamic model (such as our BAVART specification). Predictions of $\bm y_t$ are then mapped back using \autoref{eq: NS_factors}. 

The second approach follows \cite{carriero2012forecasting} and models the $M=7$ yields directly in a VAR. This approach is thus less parsimonious but also more flexible since it essentially allows for maturity-specific idiosyncrasies.

For these two approaches, we benchmark the BAVART model against several linear and non-linear competing models.  The first model we consider is a time-varying parameter (TVP) VAR with a Normal-Gamma shrinkage prior on the state innovation variances \citep[for the exact specification, see ][]{huber2020inducing}. Since the period of the zero lower bound (ZLB) is a dominant source of non-linearities in yield curve data, we estimate non-linear models that assume a dependence between the parameters of the model and the lagged short-term interest rate (in our case the one-year yield). This gives rise to the second model which is an interacted VAR (IVAR) that introduces interaction effects between the first lag of the short-term interest rate and the remaining model parameters \citep[see, e.g.,][]{caggiano2017estimating} whereas the third model is a smooth transition VAR (STVAR) which assumes that coefficients evolve slowly between two regimes \citep[see][]{gefang2009nonlinear, auerbach2012measuring}. Fourth, we consider  a Bayesian threshold VAR (TVAR) which assumes that coefficients change if the (first lag) of the short-term interest rate passes a threshold to be estimated \citep[see][]{alessandri2017financial, huber2019threshold}.  All these models are estimated using a Horseshoe prior on the VAR coefficients and benchmarked against a constant parameter NS -VAR equipped with a Minnesota prior and stochastic volatility. Moreover, we include $P=2$ lags of the endogenous variables in all models considered.

\subsection{Forecasting Results}

Before discussing the results of the forecast exercise, the question on how to compute the predictive distribution naturally arises. Producing one-step-ahead forecasts is computationally easy since conditional on the tree structures, splitting values and splitting covariates it is straightforward to compute a prediction for $\bm y_{T+1}$. More precisely (and with a slight abuse of notation), the one-step-ahead predictive distribution is given by:
\begin{equation*}
    p(\bm y_{T+1}| \bm y_{1:T}) = \int p(\bm y_{T+1}| \bm y_{1:T}, \bm \Xi) p(\bm \Xi| \bm y_{1:T})  d\bm \Xi,
\end{equation*}
where $\bm y_{1:T}$ denotes the full history of $\bm y_t$ and $\bm \Xi$ is a generic notation that summarizes all parameters and latent states (i.e. the tree structures, terminal nodes, log-volatilities etc.). This integral is solved numerically through Monte Carlo integration.

The conditional density $p(\bm y_{T+1}| \bm y_{1:T}, \bm \Xi)$ is:
\begin{equation*}
    y_{T+1}| \bm y_{1:T}, \bm \Xi \sim \mathcal{N}\left(F(\bm X_{T+1}), \tilde{\bm \Sigma}_{T+1|T}\right),
\end{equation*}
with $\tilde{\bm \Sigma}_{T+1|T}$ is a random draw obtained by using \autoref{eq: stateSV} to predict the log-volatilities and using these predictions to form $\bm H_{T+1|T}$. Higher order forecasts are then computed iteratively by first simulating from the one-step-ahead predictive distributions to obtain a prediction for $\bm y_{T+1}$. We label this draw $\bm \tilde{\bm y}_{T+1}$. $\tilde{\bm \Sigma}_{T+2|T}$ is again computed by exploiting \autoref{eq: stateSV}. This allows us to draw $\tilde{\bm y}_{T+2}$ from:
\begin{equation*}
   \tilde{\bm y}_{T+2}\sim \mathcal{N}\left(F(\tilde{\bm X}_{T+2}), \bm \Sigma_{T+2|T}\right).
\end{equation*}
We repeat this procedure until we have a prediction $\bm y_{T+o}$ with $o$ denoting the desired forecast horizon. In what follows, we will consider $o=1$ and $3$-steps-ahead forecasts.

The point (i.e. median forecasts) and density predictions are then evaluated by using relative mean squared forecast errors (MSFEs) and the continuous ranked probability score (CRPS), respectively. The MSFEs allow us to gauge the quality of point forecasts while the CRPS is used to also take into account higher order moments of the predictive distribution. All results are benchmarked against the NS-Minnesota VAR.  

\begin{figure}[t!]
\begin{minipage}[c]{0.45\textwidth}
\footnotesize \hspace{2cm}\centering{One-month-ahead}
\end{minipage}
\begin{minipage}[c]{0.45\textwidth}
\footnotesize  \hspace{4cm}\centering {One-quarter-ahead}
\end{minipage}
    \centering
    \includegraphics[scale=0.45]{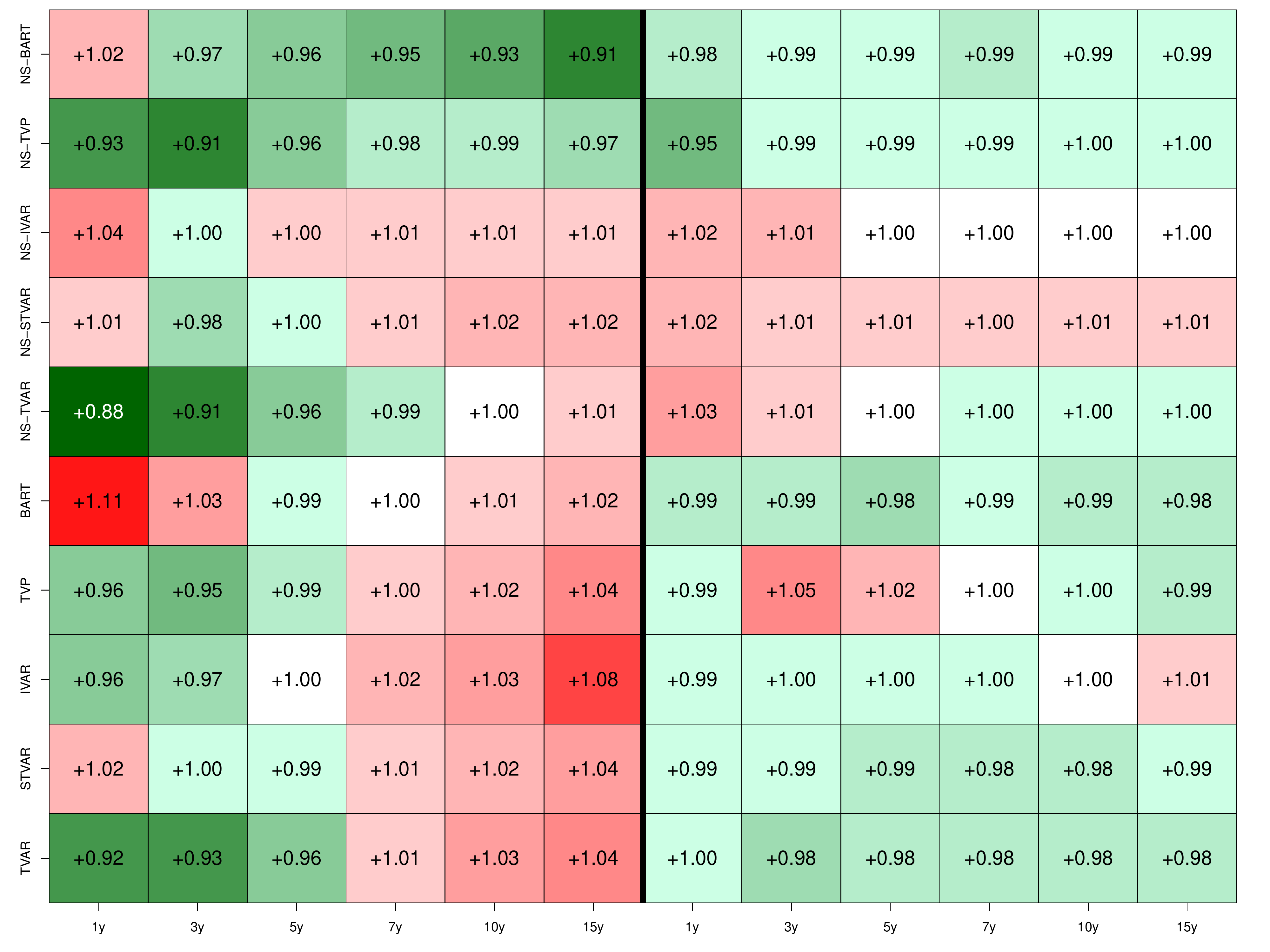}\\
        \begin{minipage}[t]{1.1\textwidth}
    \tiny \textbf{Notes}: The heatmap shows the relative mean squared forecast errors to the NS-Minnesota VAR. BART refers to the BAVART model, TVP to the time-varying parameter VAR, IVAR to the interacted VAR, STVAR to the smooth transition VAR and TVAR to the threshold VAR. The abbreviation NS refers to the Nelson-Siegel framework. The columns represent the different maturities considered.
    \end{minipage}
    \caption{Mean squared forecast errors relative to the NS-Minnesota VAR over the hold-out period: July 2006 to December 2019.}
    \label{fig:MSE}
\end{figure}

\begin{figure}[t!]
\begin{minipage}[c]{0.45\textwidth}
\footnotesize \centering{One-month-ahead}
\end{minipage}
\begin{minipage}[c]{0.45\textwidth}
\footnotesize \centering {One-quarter-ahead}
\end{minipage}
    \centering
    \includegraphics[scale=0.45]{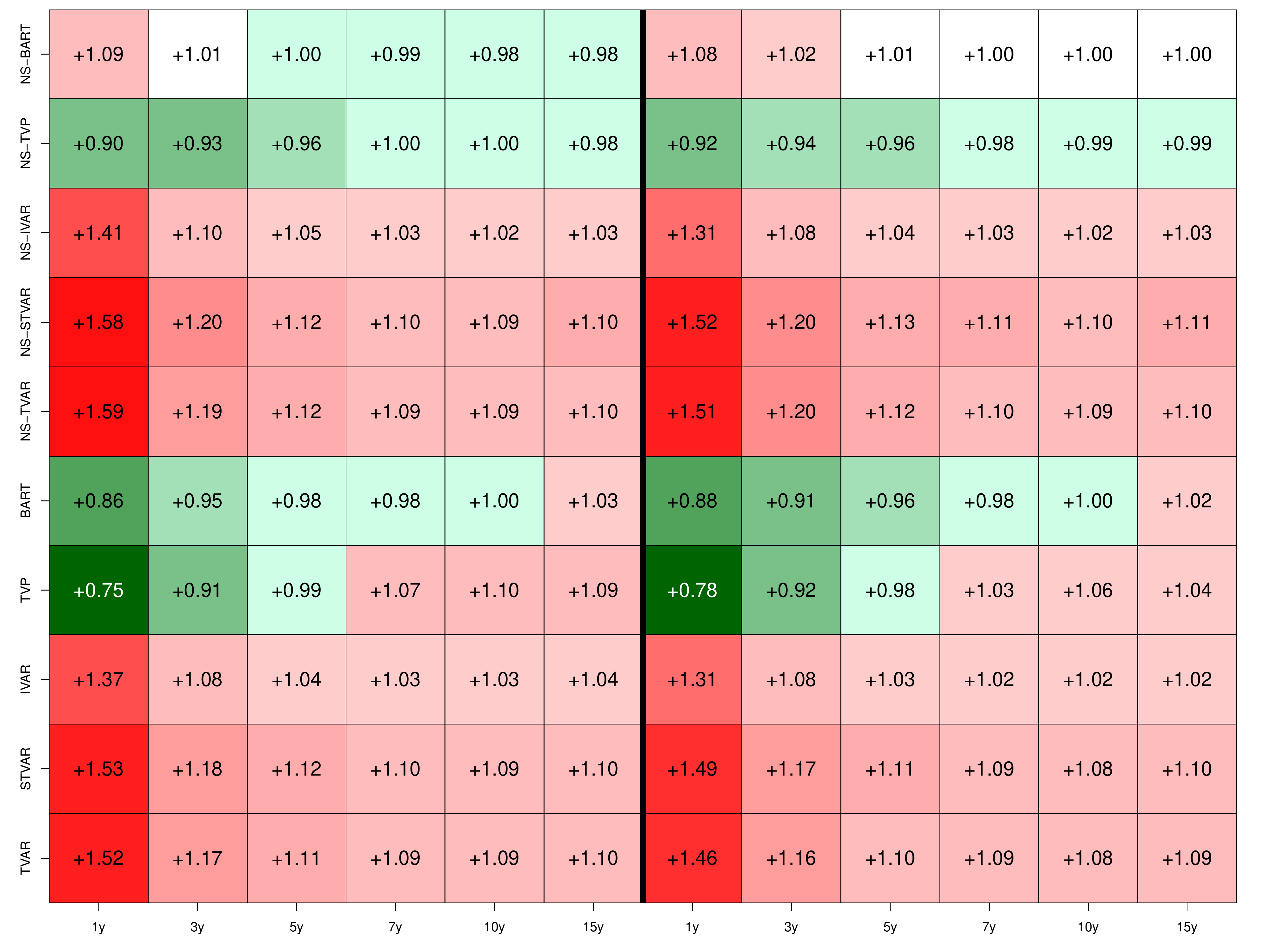}\\
            \begin{minipage}[t]{1.1\textwidth}
    \tiny \textbf{Notes}: The heatmap shows the relative mean squared forecast errors to the NS-Minnesota VAR. BART refers to the BAVART model, TVP to the time-varying parameter VAR, IVAR to the interacted VAR, STVAR to the smooth transition VAR and TVAR to the threshold VAR. The abbreviation NS refers to the Nelson-Siegel framework. The columns represent the different maturities considered.
    \end{minipage}
    \caption{Continuous ranked probability scores relative to the NS-Minnesota VAR over the hold-out period: July 2006 to December 2019.}
    \label{fig:CRPS}
\end{figure}

The findings of our forecasting exercise are summarized in heatmaps depicted in Figures \ref{fig:MSE} and \ref{fig:CRPS}.  If a given model is performing worse than the benchmark VAR, the corresponding cell will become red (i.e. a relative MSE/CRPS score exceeding unity) while if a given model outperforms the benchmark the corresponding cell will be green (with relative scores being below one).

Our forecasting horse race draws a rich picture of relative model performance. We consider models of different sizes (namely the NS variants and the ones that use the selected yields exclusively), priors, assumptions on the conditional mean, and forecast horizons.  Besides, we consider both point and density forecasts. 


Starting with one-month-ahead point forecasting accuracy, depicted in \autoref{fig:MSE}, we observe that the set of competing models improves upon the benchmark VAR. But these improvements are rather small and mostly concentrated towards the short-end (i.e. maturities shorter than seven years) of the yield curve. For this yield curve segment, there is no clear picture emerging on whether using the NS factor model or jointly modeling the different maturities improves upon the other. For some models, the NS variant yields slightly smaller MSE ratios than the unrestricted model (e.g. our BAVART model or NS-TVP), whereas for other models, unrestricted modeling of the maturities yields more precise point forecasts (e.g. the IVAR and the TVAR). 

The relative performance of our proposed BART-based model increases with the maturity. For shorter maturities, we find that BAVART performs well but in most cases is outperformed by one of the competing models (such as the NS-TVAR for one-year bonds and the NS-TVP for the three-year  bonds). When we consider bonds with maturities larger or equal than five years the performance of the NS-BART model improves appreciably. These improvements reach 7\% for 10-year government bonds and 9\% for 15-year government bonds. Interestingly, these relative MSE ratios are always smaller than the ones we observe for the unrestricted BAVART model. We conjecture that the stronger performance of BAVART for bonds with longer maturities is mainly driven by the fact that these time series display more variation, especially after the global financial crisis. This is the period where the US Federal Reserve introduced unconventional monetary policy measures such as quantitative easing to push down the slope of the yield curve.

When we focus on the one-quarter-ahead forecasts we observe a great deal of light red and green cells on the right panel of \autoref{fig:MSE}, indicating  only small gains in predictive accuracy vis-\'{a}-vis the BVAR benchmark. Most models feature MSE ratios close to unity with accuracy gains ranging from around two to five percent (in the case of the NS-TVP model and for the one-year-ahead bond yield).

Considering only point forecasts implies that we do not factor in how well a given model predicts higher order moments of the predictive distribution. We now turn to discuss the accuracy of density forecasts.
Figure \ref{fig:CRPS} shows that several of our competing models yield density forecasts which are slightly worse than the ones of the benchmark VAR for both forecast horizons. There are two exceptions to this pattern. First, we find that both versions of the BAVART and the TVP VAR yield precise density predictions for one- up to five-year yields (with the TVP VAR being better for one-year and three-year yields and the NS-BAVART outperforming all competitors for five-year yields). Especially for the short-end of the yield curve, these improvements are sizable (around 25\% for the unrestricted TVP-VAR and 14\% for the unrestricted BAVART). We conjecture that this stems from the fact that one-year yields display little variation during the period of the ZLB. In such a situation, the most flexible models in our pool (i.e. BAVART and the TVP VAR) quickly adjust both the parameters and the error variances and thus yields tighter predictive intervals centered around values close to zero.  The other models lack this flexibility since all parameters in the system either change or display no/little change. Even though the signal variable is the short-term interest rate this potentially is a limitation and thus could negatively impact density forecasting accuracy. 

Second, when we focus on three-months-ahead forecasts a similar picture emerges. The  unrestricted BAVART  and the TVP model work well at the short end of the yield (with sizable gains for one- and three-year yields in both cases). For higher maturities, and in contrast to the results shown in Figure \ref{fig:MSE}, we find relative CRPSs close to one, indicating that all models included have a hard time beating the benchmark VAR with SV.

To sum up, our forecasting exercise shows that the BAVART model works remarkably well for predicting the US term structure of interest rates. When we focus on point forecasts, the predictive gains of the BAVART model increase with the maturity of the bond. For density predictions, this story is reversed, indicating that both variants of the BAVART work well at the short-end of the yield curve.

\section{The Effects of Macroeconomic Uncertainty on the Eurozone}\label{sec: IRF_EA}
We now turn to our application based on Eurozone data. The next sub-section provides a brief introduction to the dataset used while  Sub-section \ref{sec:insample} shows some in-sample results. Sub-section \ref{sec: IRFs} deals with the question of how macroeconomic uncertainty impacts the Eurozone economy with a particular emphasis on the zero lower bound.
\subsection{Data overview}
In the second application, we apply our BAVART model to analyze the effects of uncertainty shocks on macroeconomic outcomes. Several recent papers analyze this question using US datasets \citep{bloom2009impact, jurado2015measuring, caggiano2017estimating, Ramey2017, carriero2018measuring}. Instead of focusing on US data, we apply our approach to monthly Eurozone data that ranges from January 1999 to January 2019. We opt for this dataset because the Eurozone economy underwent structural changes over this estimation period, multiple recessions and in general a challenging environment since the time series we consider are rather short.

The dataset we use is a selection of time series from the popular Euro Area Real Time Database (EA-RTD).  We consider a medium-scale model that includes six macroeconomic and financial time series. These six time series are augmented by an uncertainty index (abbreviated as {UNC}) which is estimated using the approach outlined in \cite{jurado2015measuring}. Apart from the uncertainty indicator, we include the (log) level Dow Jones Euro Stoxx 50 price index (DJE50), HICP inflation (C\_OV),  unemployment rate (UNETO),  3-month Euribor (EUR3M), industrial production (XCONS), and the yield on Euro area 10-year benchmark government bonds (10Y).  We include one lag of the endogenous variables.\footnote{Using a single lag allows for simple inspection of several features of the model. Including more lags is straightforward but lags larger than one only rarely show up in the splitting rules and the impulse responses look very similar to the ones reported in the main body of the text.}

\subsection{In-sample results}\label{sec:insample}
In this section, we illustrate key features of our BART-based approach using a subset of the dataset discussed in the previous section.  Variable importance is gauged by considering the posterior median of the number of times a given quantity shows up in a splitting rule. These frequencies are shown in Table \ref{tab: varIMPORTANCE}. The columns refer to the  different equations  and the rows to the corresponding covariates. 

\begin{table}[ht]
\centering
\begin{tabular}{lrrrrrrr}
  \toprule
 & UNC${t}$ & DJE50$_t$ & EUR3M$_t$ &  C\_OV$_t$ &   UNETO$_t$ & XCONS$_t$ &   10Y$_t$ \\ 
  \midrule
UNC$_{t-1}$                & 52 & 44 & 44 & 38 & 41 & 45 & 44 \\
  DJE50$_{t-1}$            & 46 & 56 & 44 & 32 & 39 & 43 & 47 \\
  EUR3M$_{t-1}$            & 49 & 50 & 62 & 55 & 40 & 47 & 45 \\
  C\_OV$_{t-1}$            & 47 & 48 & 48 & 60 & 43 & 50 & 50 \\
  UNETO$_{t-1}$            & 47 & 46 & 27 & 38 & 60 & 49 & 47 \\
  XCONS$_{t-1}$            & 45 & 43 & 43 & 30 & 45 & 51 & 47 \\
  10Y$_{t-1}$              & 43 & 42 & 33 & 37 & 41 & 43 & 51 \\
   \bottomrule
\end{tabular}
\caption{Number of times a given covariate shows up across splitting rules in a medium-scale VAR with one lag.} \label{tab: varIMPORTANCE}
\end{table}

A simple inspection of the main diagonal elements of the table reveals that the first, own lag of a given variable within the corresponding equation shows up frequently. This finding holds for all equations and resembles key results of the literature on Bayesian VARs which states that the AR(1) term explains most variation in $\bm y_t$. However, here it is worth stressing that our model is far from sparse, and also the lags of other variables seem to play a role in determining the dynamics of a given endogenous variable. The lags of other quantities in a given equation are also often included as splitting variables, indicating that we can not tell a simple story about some few elements in $\bm x_t$ exclusively shaping the dynamics of $\bm y_t$.

\subsubsection{Impulse Responses to an Uncertainty Shock and the Role of the Zero Lower Bound}\label{sec: IRFs}
In this section, we illustrate how the BAVART model can be used to carry out structural inference. There is a broad body of literature dealing with the macroeconomic effects of uncertainty using non-linear models \citep{ferrara2018macroeconomic, mumtaz2018changing,alessandri2019financial,cuaresma2020fragility, paccagnini2020asymmetric, caggiano2017uncertainty}. These papers all rely on parametric approaches and make rather strong assumptions on the nature of non-linearities. Our BAVART model, by contrast, introduces almost no restrictions and thus remains agnostic on the specific form of non-linearities in the transmission mechanisms.

Because of the highly non-linear nature of our model, we resort to  generalized impulse response (GIRF) functions \citep[see][]{koop1996impulse}. These impulse responses are computed as follows. Let $\bm \epsilon_t$ denote the $M$ structural shocks and $\bm s_j$ denotes a $M\times 1$ selection vector that equals $1$ in the $j^{th}$ position. Hence, $\epsilon_{jt} = \bm s'_j \bm \epsilon_t$ yields the $j^{th}$ structural shock. The GIRF to the $j^{th}$ structural shock is then defined as the difference between the forecast which assumes $\epsilon_{jt}=1$ (while setting the other shocks to zero) and the unconditional forecast (i.e. with $\bm \epsilon = \bm 0_M$) for a given forecast horizon.  The posterior distributions of the GIRFs are computed by repeating this procedure during MCMC sampling. The structural shock is computed by setting $\bm H_t$ equal to the unconditional mean, i.e. $\bm H_t = \text{diag}(e^{c_1}, \dots, e^{c_M})$ and assuming that the uncertainty factor is ordered first.

To assess whether BART uncovers implicit non-linear relations consistent with the literature quoted above, we consider two  experiments. The first assumes that the economy is hit by a one standard deviation uncertainty shock. The dynamic reactions of $\bm y_t$ are left unrestricted. This implies that the central bank is able to react to increases in uncertainty by lowering interest rates.

The second experiment asks how the reactions in $\bm y_t$ change if the economy is stuck at the zero lower bound (ZLB) and the central bank can not use conventional monetary policy tools. This second experiment is interesting because our BAVART model can use the information that the short-term interest rate is zero so that different branches of a given tree are effectively ruled out.  For instance, consider a simple tree which has a root node with a decision rule that splits the observations according to whether the  short-term interest rate is below one percent or greater than one percent. If we introduce the restriction that the interest rate is zero, the second branch of the tree (the above one percent interest rate branch) will play no role in constructing the impulse responses.  Hence, only different configurations of the covariate space which are consistent with interest rates close to zero are being considered. The restriction that the short-term interest rate is at the ZLB is easily incorporated by zeroing out the forecast of the short rate. 

\begin{figure}[!tbp]
    \centering
    \includegraphics[scale=0.7]{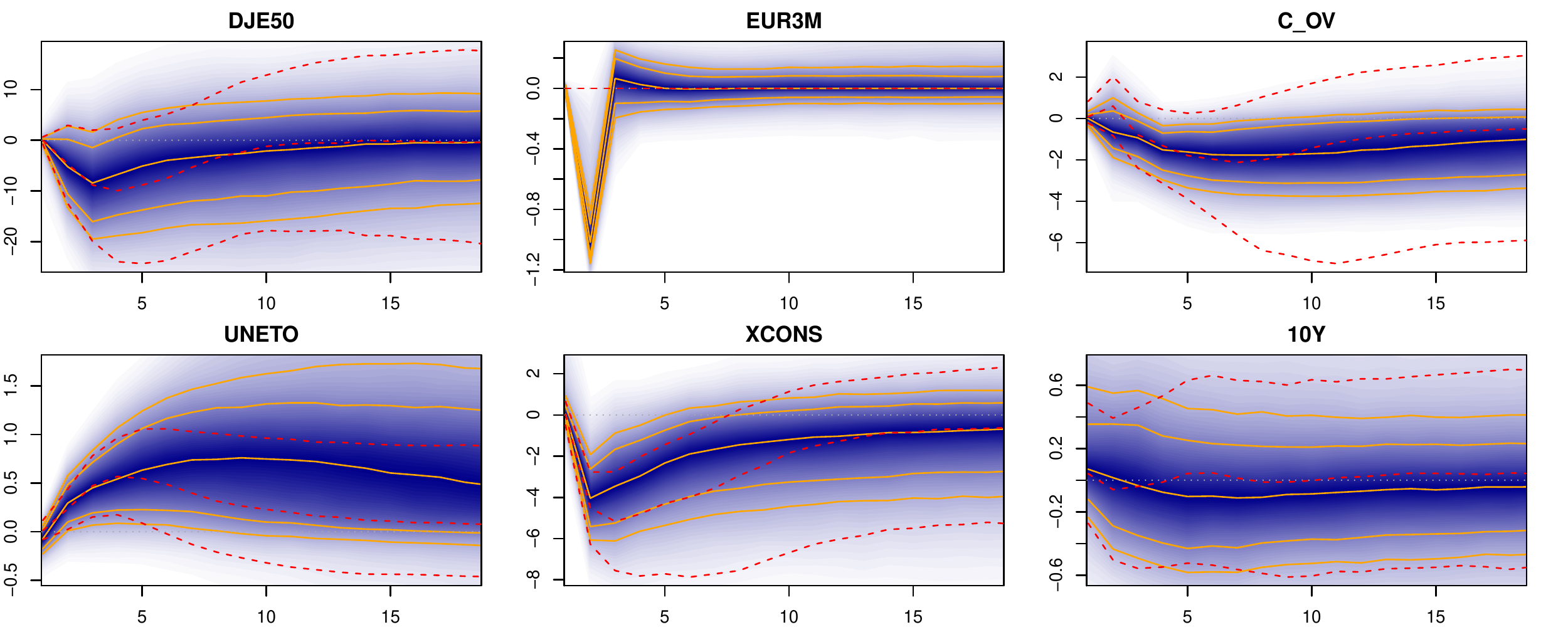}\\
    \begin{minipage}[t]{1\textwidth}
    \tiny \textbf{Notes}: The shaded blue area represents the marginal posterior distributions of impulse responses while the solid orange lines represent the $16 (25)$th and $84 (25)$th percentiles of the posterior distribution. The dotted red lines represent the $16$th and $84$th credible intervals of the impulse responses which restrict the short-term interest rate reaction to zero. The dotted dark gray line denotes zero.
    \end{minipage}
\caption{Dynamic responses to a one-standard deviation uncertainty shock. The dashed gray lines represent $16$th and $84$th posterior credible intervals.} \label{fig: IRFs}
\end{figure}

Figure \ref{fig: IRFs} presents the dynamic responses to a one standard deviation uncertainty shock. The blue shaded area is the marginal posterior distribution of the IRFs, with solid orange lines denoting the $16 (25)$th and $75 (84)$th percentiles.  The dotted red lines represent the $16$th and $84$th credible intervals of the impulse responses based on zeroing out the reaction of the short-term interest rate.

In general, we find that the IRFs closely resemble the ones reported in the literature \citep[see, among many others,][]{bloom2009impact, jurado2015measuring, mumtaz2018changing}. Focusing on the reactions of stock markets (DJE50), there is some limited evidence that equity prices decline. If interest rates are stuck at the ZLB, these declines are somewhat more pronounced and appear to peak slightly later. 

Considering the dynamic responses of short-term interest rates (EUR3M) reveals that after a month, short rates decrease appreciably.  This decline in interest rates peaks after about two months and reaches around 100 basis points. In economic terms, the negative reaction of short-term interest rates is likely induced by expansionary monetary policy measures undertaken by the central bank. Notice that in the presence of the ZLB, we assume no interest rate reaction at the short-end of the yield curve.

Turning to inflation reactions (C\_OV) suggests that prices tend to fall after about six months. This decline in inflation is consistent with a negative demand channel which implies that firms lower prices in response to a decline in aggregate demand \citep[see][]{bloom2009impact}. When the economy is stuck at the ZLB, we observe somewhat stronger but insignificant inflation reactions. These responses are consistent with \cite{caggiano2017estimating} who also find more pronounced  but insignificant inflation responses if short rates hit the ZLB.

The unemployment rate (UNETO), with a lag of around two to three months, increases and remains elevated for several months before turning insignificant. This increase in the unemployment rate is consistent with \cite{jurado2015measuring} and \cite{carriero2018measuring} who, for US data, find similar unemployment responses.  One interesting finding is that the unemployment responses in the presence of the ZLB are similar in magnitude but tend to peter out faster and thus are more short-lived than the ones if interest rates are allowed to react freely. It is noteworthy that the impulse responses are very similar over the first four to five months and then depart from each other. 

When we consider industrial production (XCONS) we observe a sluggish decline which peaks at around minus five percent in the second month and then, after around six months the reactions of real activity turn insignificant. Interestingly, we do not observe a rebound in real activity arising from a ''wait-and-see" mechanism reported in, e.g., \cite{bloom2009impact}. \cite{gieseck2016impact}, for Eurozone data, find similar responses for GDP which are also rather short-lived and do not display a substantial real activity overshoot. When we assume that the ZLB is binding, the reactions of industrial production become more pronounced, which is consistent with other findings who report that if interest rates hit zero, real activity reacts stronger to uncertainty shocks \citep{caggiano2017uncertainty}. Finally, long-term interest rates display no statistically significant reaction throughout the impulse response horizon.

To sum up, a key take away from this exercise is that our BAVART model is capable of producing meaningful impulse responses which are consistent with the literature. Without introducing any assumptions on the specific form of non-linearities but in the presence of the ZLB, our BAVART specification yields impulse responses that are consistent with papers that assess how the uncertainty -- real activity nexus changes if the central bank is constrained by the ZLB.

\section{Conclusions}
\label{sec_Conclusions}
VAR models  assume that the lagged dependent variables influence the contemporaneous values in a linear fashion. In this paper, we relax this assumption by blending the literature on BART models and VARs. The resulting BAVART model can handle arbitrary non-linear relations between the endogenous and the exogenous variables. Our proposed model is, moreover, capable of handling stochastic volatility in the shocks. As opposed to existing models which make strong assumptions on the nature of non-linearities, our model remains agnostic and allows to estimate these forms in a data-based manner. To make the model operational, we briefly discuss Bayesian estimation but also show  how to compute multi-step-ahead forecasts and generalized impulse responses.

We illustrate our approach using two topical applications. In the first application  we apply the model to the US term structure of interest rates. Using several linear and non-linear competing models and different ways of modeling the yield curve, we show that our BAVART model yields precise point and density forecasts. The point forecasting accuracy differences become larger with the maturity of a given bond. An opposite picture emerges when the full predictive distribution is used: in that case, forecast gains can be mostly found at the short-end of the yield curve.

The second application deals with the effects of uncertainty on the Eurozone economy. To investigate the role of the ZLB on interest rates, we consider impulse responses which restrict interest rate reactions to zero and compare these to their unrestricted counterpart. The findings indicate that uncertainty has a detrimental effect on macroeconomic outcomes.  Unemployment increases, prices fall, stock markets  decline and industrial production drops markedly. In general, these reactions are similar to other findings in the literature which mainly focus on US data. When we assume that interest rates are stuck at the ZLB, real activity responses become somewhat more elevated. Especially for output, we observe much stronger responses if the central bank is not able to react adequately.

There are many possible avenues for further research and possible applications of our model. For instance, the model can be used to track asymmetries in the transmission of economic shocks. Or it could be applied to high-frequency financial data such as daily stock returns and then combined with a heavy-tailed error distribution to produce precise density forecasts. From an econometric perspective, BART could be used to model time-variation in regression coefficients and thus generalize TVP regressions which assume a random walk evolution on the latent states.

\bibliographystyle{apalike}
\bibliography{BART_Biblio}

\newpage

%
\renewcommand{\thesection}{A.\arabic{section} }
\renewcommand{\theequation}{A.\arabic{equation}}
\renewcommand{\thefigure}{A.\arabic{figure}}
\renewcommand{\thetable}{A.\arabic{table}}
\setcounter{table}{0}
\setcounter{figure}{0}
\setcounter{equation}{0}
\appendix{}

%
\section{Posterior Approximation of the Trees}\label{App_Sec_Posterior}
In this appendix, we provide details on the MCMC algorithm used to simulate from the joint posterior distribution of the trees and the terminal node parameters. The remaining quantities take well known forms and are very easy to simulate using textbook results for the Gaussian linear regression model.

As stated in \autoref{sec_prior} and following \cite{Carriero2019}, \cite{Koop2019} and \cite{Kastner2020}, we rely on a structural representation of the model that entails equation-by-equation estimation.

We simulate from the conditional posterior of $(\mathcal{T}_{jk},\bm{m}_{jk})$ by conditioning on all trees except  the $k^{th}$ one. This is achieved by using the Bayesian backfitting strategy discussed in \cite{Chipman2010}. More precisely,  the likelihood function depends on $(\mathcal{T}_{jk},\bm{m}_{jk})$ through the partial residuals
\begin{equation*}
\bm R_{jk} = \bm{y}_{\bullet j} - \sum_{l=1}^{j-1} a_{jl} \bm \varepsilon_{\bullet l} - \sum_{s \neq k} g_{j s}(\bm X| \mathcal{T}_{j s}, \bm m_{j s}).
\end{equation*}
The algorithm proposed in \cite{Chipman2010} then proceeds by integrating out  $\bm{m}_{jk}$ and then drawing $\mathcal{T}_{jk}$ with Metropolis-Hastings (MH) algorithm detailed in \cite{Chipman1998}. 

This step is implemented by generating a candidate tree $\mathcal{T}_{jk}^{\ast}$ from a proposal  distribution $q(\mathcal{T}_{jk},\mathcal{T}_{jk}^{\ast})$ and then  accept the proposed value with probability:
\begin{equation*}
\alpha(\mathcal{T}_{jk},\mathcal{T}_{jk}^{\ast}) = \min{\left\{1,\frac{q(\mathcal{T}_{jk}^{\ast},\mathcal{T}_{jk})}{q(\mathcal{T}_{jk},\mathcal{T}_{jk}^{\ast})} \frac{p(\bm R_{ jk}|\bm{X},\mathcal{T}_{jk}^{\ast},M_j)}{p(\bm R_{jk}|\bm{X},\mathcal{T}_{jk},M_j)} \frac{p(\mathcal{T}_{jk}^{\ast})}{p(\mathcal{T}_{jk})} \right\}}.
\end{equation*}
The first term is the ratio of the probability of how the previous tree moves to the new tree against the probability of how the new tree moves to the previous one. The second term is the likelihood ratio of the new tree against the previous tree and the last term denotes the likelihood ratio of the new against the previous tree under the prior distribution.

The proposal distribution $q(\mathcal{T}_{jk},\mathcal{T}_{jk}^{\ast})$ features four local steps \citep{Chipman1998}. The first  step grows the tree by splitting a node into two different nodes. This step is chosen with probability 0.25 The second step combines two non-terminal nodes into a single terminal node. We select this step with probability 0.25 as well.  The third step swaps splitting rules between two terminal nodes (with probability 0.4) and the final step changes the splitting rule of a single non-terminal node (with probability 0.1).

After sampling the tree structure we can easily simulate from the posterior distribution of the terminal nodes. Since the prior is conjugate the resulting posterior will also be a Gaussian distribution that takes a standard form and does not explicitly depend on the tree structure but only on the subset of elements in the residuals allocated to a specific terminal node.

\clearpage

\renewcommand{\thesection}{B.\arabic{section}}
\renewcommand{\theequation}{B.\arabic{equation}}
\renewcommand{\thefigure}{B.\arabic{figure}}
\renewcommand{\thetable}{B.\arabic{table}}
\setcounter{table}{0}
\setcounter{figure}{0}
\setcounter{equation}{0}

\end{document}